\documentclass[aps,prb,reprint,superscriptaddress,showpacs]{revtex4-2}
\usepackage{graphicx}
\usepackage{bm,amsmath,amssymb,mathrsfs,dcolumn,}
\usepackage[colorlinks=true, linkcolor=blue, citecolor=blue, urlcolor=blue, linktoc=page, bookmarks=false]{hyperref} 
\usepackage[usenames]{color}
\hypersetup{pdfborder=1 1 1,colorlinks=true,citecolor=blue,linkcolor=blue}
\usepackage{epstopdf}
\usepackage{subfigure}
\usepackage{units}
\usepackage{breakurl}
\usepackage{multirow}
\usepackage{booktabs}

\UseRawInputEncoding
\begin{document}


\title{Comparison of spin-wave transmission in parallel and antiparallel magnetic configurations}

\author{Y. W. Xing}
\thanks{These two authors contributed equally to this work.}
\affiliation{Beijing National Laboratory for Condensed Matter Physics, Institute of Physics, University of Chinese Academy of Sciences, Chinese Academy of Sciences, Beijing 100190, China}
\affiliation{Center of Materials Science and Optoelectronics Engineering, University of Chinese Academy of Sciences, Beijing 100049, China}

\author{Z. R. Yan}
\thanks{These two authors contributed equally to this work.}
\affiliation{Beijing National Laboratory for Condensed Matter Physics, Institute of Physics, University of Chinese Academy of Sciences, Chinese Academy of Sciences, Beijing 100190, China}

\author{X. F. Han}
\email{xfhan@iphy.ac.cn}
\affiliation{Beijing National Laboratory for Condensed Matter Physics, Institute of Physics, University of Chinese Academy of Sciences, Chinese Academy of Sciences, Beijing 100190, China}
\affiliation{Center of Materials Science and Optoelectronics Engineering, University of Chinese Academy of Sciences, Beijing 100049, China}
\affiliation{Songshan Lake Materials Laboratory, Dongguan, Guangdong 523808, China}



\date{\today}
\begin{abstract}
Parallel (P) and antiparallel (AP) configurations are widely applied in magnetic heterostructures and have significant impacts on the spin-wave transmission in magnonic devices. In the present study, a theoretical investigation was conducted into the transmission of exchange-dominated spin waves with nanoscale wavelengths in a type of heterostructure including two magnetic media, of which the magnetization state can be set to the P (AP) configuration by ferromagnetic (antiferromagnetic) interfacial exchange coupling (IEC). The boundary conditions in P and AP cases were derived, by which the transmission and reflection coefficients of spin waves were analytically given and numerically calculated. In the P configuration, a critical angle $\theta_{\textrm{c}}$ always exists and has a significant influence on the transmission. Spin waves are refracted and reflected when the incident angle $\theta_{\textrm{i}}$ is smaller than the critical angle ($\theta_{\textrm{i}} < \theta_{\textrm{c}}$), while total reflection occurs as $\theta_{\textrm{i}} \geq \theta_{\textrm{c}}$. In the AP configuration, the spin-wave polarizations of medium 1 and 2 are inverse, that is, right-handed (RH) and left-handed (LH), leading to the total reflection being independent of $\theta_{\textrm{i}}$. As demonstrated by the difference in spin-wave transmission properties between the P ($\theta_{\textrm{i}} < \theta_{\textrm{c}}$) and AP cases, there is a polarization-dependent scattering. However, as $\theta_{\textrm{i}}$ exceeds $\theta_{\textrm{c}}$, the P ($\theta_{\textrm{i}} > \theta_{\textrm{c}}$) case exhibits similarities with the AP case, where the transmitted waves are found to be evanescent in medium 2 and their decay lengths are investigated. In both the P ($\theta_{\textrm{i}} > \theta_{\textrm{c}}$) and AP cases, the Goos-H\"{a}nchen (GH) shift of the total reflection waves are calculated and shown as a function of the frequency and incident angle. The relationship between the decay lengths and GH shifts is also explored. Further, as the number of media exceeds two, spin waves are scattered by multiple interfaces, resulting in the resonant transmission effect in the P ($\theta_{\textrm{i}} < \theta_{\textrm{c}}$) case. At the same time, there is a tunnelling effect and a resonant tunnelling effect in the P ($\theta_{\textrm{i}} > \theta_{\textrm{c}}$) and AP cases, which are attributed to the evanescent waves. The influences of the IEC strength on all of the aforementioned findings are investigated in detail. The present study provides a comprehensive guide for the transmission of spin waves in the magnetic systems with either P or AP configuration and is helpful for the design of future magnonic devices.

\end{abstract}

\maketitle
\section{Introduction}
In the field of traditional spintronics, the relative orientation of magnetizations has a significant impact on the transport properties of electrons, such as the different magnetoresistance between parallel (P) and antiparallel (AP) configurations in the spin valves \cite{RN2,RN3} and magnetic tunneling junctions (MTJs) \cite{RN1,RN9,RN10}. In 1975, Julli¨¨re studied the conductance of Fe/Ge/Co junctions at 4.2 K in P and AP states \cite{RN1}. The difference in conductance between the two states demonstrated that the scattering of electrons is spin-dependent, which is referred to as tunneling magnetoresistance (TMR).

The application of such magnetoresistance effect in technology is difficult due to the necessity of low temperature. In the late 1980s, giant magnetoresistance (GMR) was discovered by Fert \cite{RN2} and Gr\"{u}nberg \cite{RN3}, which greatly facilitated further researches and applications, and is generally regarded as the beginning of spintronics \cite{RN4}. The ferromagnetic/nonmagnetic/ferromagnetic sandwiches are the key structures for realizing GMR, of which the resistance exhibits a significant change between the P and AP states at both low and room temperatures. Subsequently, a type of spin valve was proposed in respect of the antiferromagnetic pinning \cite{RN5,RN6,RN7,RN8}, which is extensively used in magnetic read heads and sensors. In 1995, the room-temperature TMR was discovered by Miyazaki \cite{RN9} and Moodera \cite{RN10} in MTJs with the $\textrm{Al}_{2}\textrm{O}_{3}$ insulating barrier. Such discovery further promoted the development of spintronics.

Inspired by the effects of GMR and TMR, studies have been conducted regarding the transmission properties of other particles or quasi-particles in P and AP configurations, such as magnons. As the elementary excitation of the magnetic system, spin waves, or magnons, are regarded as potential information carriers. The research field of magnons is called magnonics \cite{RN10.5,RN11,RN12,RN13,RN14}, in which the kernel is to manipulate the magnon transmission by various designs \cite{RN15,RN16,RN17,RN18,RN19,RN20,RN21,RN22,RN23,RN24,RN25,RN26}. Among these researches, Wu \textit{et} \textit{al}. \cite{RN19} fabricated a new type of device, YIG/Au/YIG, which was called the magnon valve and drew widespread attention. Referred to as the magnon valve effect (MVE), magnon currents can pass through the magnon valve in the P state, but are blocked in the AP state. After, the magnon junction, YIG/NiO/YIG, was proposed \cite{RN20} as the counterpart of MTJs due to the insulating barrier. The aforementioned studies demonstrated the significant influence of magnetic configurations on the spin-wave transmission.

Notably, spin waves are thermally excited by the spin Seebeck effect (SSE) in magnon valve \cite{RN19} and magnon junction \cite{RN20} experiments, thus the coherence of spin waves is ignored. In fact, the coherence is significant for spin-wave transmission because certain phenomena can occur, such as refraction \cite{RN27,RN28,RN29,RN30,RN31,RN32,RN33}, skin effect \cite{RN34}, total reflection and decay \cite{RN74}, Goos-H\"{a}nchen (GH) shift \cite{RN35,RN36,RN37,RN38}, tunneling \cite{RN39}, resonant tunneling \cite{RN40,RN41} and resonant transmission \cite{RN42,RN43,RN44,RN45}. Among these works, the present authors \cite{RN34} and Poimanov \textit{et} \textit{al}. \cite{RN74} have investigated the transmission properties of spin waves in the AP configuration and found the evanescent waves induced by the inverse polarization, which were significantly different with the P ($\theta_{\textrm{i}} < \theta_{\textrm{c}}$) case. However, few studies were reported on evanescent waves in the P configuration with $\theta_{\textrm{i}} > \theta_{\textrm{c}}$. Hence, the systematic research on the transmission of coherent spin waves in P and AP magnetic configurations is essential.

In the present study, the focus is on the coherent exchange-dominated spin waves (wavelength $\lambda <$ 100 nm) \cite{RN46,RN47}, which are significant for the nanoscale magnonic devices. The main system of the present research is comprised of two magnetic media and has two states, including the P and AP configurations, corresponding to the ferromagnetic and antiferromagnetic interfacial exchange coupling (IEC), respectively. The boundary conditions at the interface between two media were analytically deduced in both the P and AP cases. Based on the boundary conditions, the expressions of transmission and reflection coefficients were obtained. The computed results show obvious differences between the P and AP states. In the P state, the critical angle $\theta_{\textrm{c}}$ is found to always exist. When the incident angle $\theta_{\textrm{i}}$ is smaller than the critical angle $\theta_{\textrm{c}}$, spin waves are refracted and reflected. When $\theta_{\textrm{i}} \geq \theta_{\textrm{c}}$, spin waves are all reflected. In the AP state, there is no critical angle and total reflection invariably occurs. Such findings can be attributed to the inverse spin-wave polarizations of medium 1 and 2, that is, right-handed (RH) and left-handed (LH). Further, in the cases of total reflection in both the P and AP states, the spin waves are found to penetrate into medium 2 in the form of evanescent waves. The decay lengths of such evanescent waves are analytically and numerically studied. The relationship between the decay lengths and GH shifts is also explored. Additionally, the spin waves propagating in multiple media are investigated, in which the phenomena of resonant transmission, tunnelling and resonant tunnelling occur.
\section{Analytical model}
As shown in Figure 1(a), a system of two magnetic media is considered with $\textbf{m}_{\textrm{n}}$ (n = 1 or 2) being the unit magnetization vector. The magnetic media can be either ferromagnetic or ferrimagnetic in the model. The type of IEC between medium 1 and 2 can be ferromagnetic and antiferromagnetic \cite{RN48}, corresponding to P ($A_{12} > 0$) and AP ($A_{12} < 0$) states with $A_{12}$ being the constant of IEC. The dynamics of $\textbf{m}_{\textrm{n}}$ is governed by the Landau-Lifshitz-Gilbert (LLG) equation \cite{RN49,RN50}
\begin{equation}
\label{eq.1}
\frac{\partial\mathbf{m}_\textrm{n}}{\partial t}=-\gamma\mu_0\mathbf{m}_\textrm{n}\times\mathbf{H}_\textrm{n}^{\textrm{eff}}+\alpha\mathbf{m}_\textrm{n}\times\frac{\partial\mathbf{m}_\textrm{n}}{\partial t},
\end{equation}
where $\gamma = 1.76 \times 10^{11}$ rad/(s T) is the gyromagnetic ratio, $\mu_{0}$ is the vacuum permeability, $\alpha$ is the Gilbert damping coefficient, $\mu_{0}\textbf{\textrm{H}}_{\textrm{n}}^{\textrm{eff}}=\frac{2A_{\textrm{n}}}{M_{\textrm{n}}}\nabla^{2}\textbf{m}_{\textrm{n}}+\sigma_{\textrm{n}}\frac{2K_{\textrm{n}}}{M_{\textrm{n}}}\textbf{e}_{\textrm{z}}$ is the effective field with the saturation magnetization $M_{\textrm{n}}$, exchange constant $A_{\textrm{n}}$ and uniaxial magnetic anisotropy constant $K_{\textrm{n}}$, and $\sigma_{\textrm{n}}$ is the orientation factor of $\textbf{m}_{\textrm{n}}$. As shown in Figure 1(b) and (c), if $\sigma_{\textrm{n}}$ = +1 (-1), $\textbf{m}_{\textrm{n}}$ is parallel (antiparallel) to $+z$ axis, representing that the spin-wave polarization of the medium is RH (LH). The internal torque is $\bm{\tau}_{\textrm{n}}^{\textrm{In}}=-\gamma\mu_{0}\textbf{m}_{\textrm{n}}\times\textbf{\textrm{H}}_{\textrm{n}}^{\textrm{eff}}$. For an independent spin-up or spin-down medium, the internal spin waves are right-handed circularly polarized (RHCP) or left-handed circularly polarized (LHCP). However, due to the IEC, the situation gets complicated for a system of two coupled media.

\begin{figure}[t]
  \centering
  \includegraphics[width=7cm]{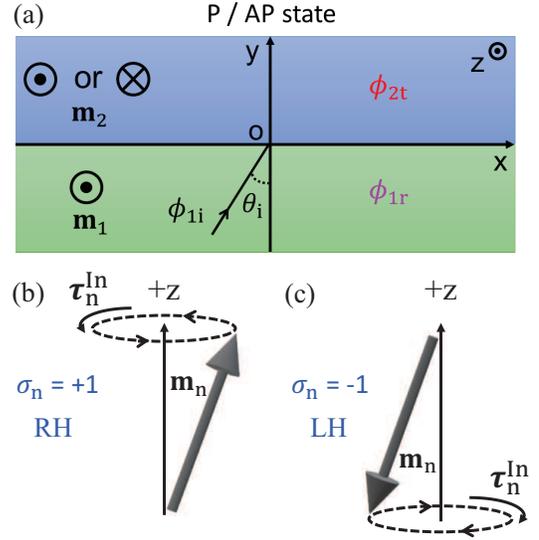}
  \caption{(a) The schematic diagram of the two-medium system in the P or AP state. \textbf{m}$_{1}$ is in the $+z$ direction, while \textbf{m}$_{2}$ is along $+z$ ($-z$) in the P (AP) state. $\phi_{\textrm{1i}}$ and $\phi_{\textrm{1r}}$ represent the incident and reflected waves in medium 1, while $\phi_{\textrm{2t}}$ denotes the transmitted waves in medium 2. $\theta_{\textrm{i}}$ is the incident angle and the incident point is the origin of coordinates. (b) and (c) show the RH and LH spin-wave polarizations, corresponding to $\sigma_{\textrm{n}} = +1$ and $-1$, respectively.}\label{Fig1}
\end{figure}

In the P (AP) state of the two-medium system, $\sigma_{1}$ = $\sigma_{2}$ = +1 ($\sigma_{1}$ = +1 and $\sigma_{2}$ = $-1$), and the spin-wave polarizations of medium 1 and 2 are the same (inverse). In order to solve the problem of spin-wave propagation in the system, the boundary conditions at the interface ($y = 0$) need to be derived. A disk is considered that covers the part from $y = -\varepsilon$ to $y = 0$ with the thickness $\varepsilon \ll 1$ and the cross section area being $\xi$. Integrating Eq. (1) over the volume of the disk and neglecting the damping $\alpha$, one can get \cite{RN43,RN51}
\begin{equation}
\label{eq.2}
I_{\textrm{ex}}+I_{\textrm{IEC}}=0,
\end{equation}
where
\begin{equation}
\label{eq.3}
\begin{split}
I_{\textrm{ex}}&=\int{\mathbf{m}_1\times\mathbf{H}_{\textrm{ex},1}}\textrm{dV}\\
&=\frac{2A_1}{\mu_0M_1}\int{\mathbf{m}_1\times\nabla^2\mathbf{m}_1}\textrm{dV}\\
&=-\xi\frac{2A_1}{\mu_0M_1}\mathbf{m}_1\times\frac{\partial\mathbf{m}_1}{\partial y}
\end{split}
\end{equation}
and
\begin{equation}
\label{eq.4}
\begin{split}
I_{\textrm{IEC}}&=\int{\mathbf{m}_1\times\mathbf{H}_{\textrm{IEC},1}}\textrm{dV}\\
&=\frac{2A_{12}}{\mu_0M_1\varepsilon}\int{\mathbf{m}_1\times\mathbf{m}_2}\textrm{dV}\\
&=\xi\frac{2A_{12}}{\mu_0M_1}\mathbf{m}_1\times\mathbf{m}_2
\end{split}
\end{equation}
are the only terms which can not be neglected. Uniting Eq. (2)-(4), it leads to
\begin{equation}
\label{eq.5}
A_1\mathbf{m}_1\times\frac{\partial\mathbf{m}_1}{\partial y}={A_{12}\mathbf{m}}_1\times\mathbf{m}_2.
\end{equation}
Similarly, another disk can be considered that covers the part from $y = 0$ to $y = \varepsilon$ with the thickness $\varepsilon \ll 1$ and the cross section area being $\xi$. Integrating Eq. (1) over the volume of this disk and neglecting the damping, the following equations can be obtained \cite{RN43}:
\begin{equation}
\label{eq.6}
I_{\textrm{ex}}^\prime+I_{\textrm{IEC}}^\prime=0,
\end{equation}
where
\begin{equation}
\label{eq.7}
\begin{split}
I_{\textrm{ex}}^\prime &=\int{\mathbf{m}_2\times\mathbf{H}_{\textrm{ex},2}}\textrm{dV}\\
&=\frac{2A_2}{\mu_0M_2}\int{\mathbf{m}_2\times\nabla^2\mathbf{m}_2}\textrm{dV}\\
&=\xi\frac{2A_2}{\mu_0M_2}\mathbf{m}_2\times\frac{\partial\mathbf{m}_2}{\partial y}
\end{split}
\end{equation}
and
\begin{equation}
\label{eq.8}
\begin{split}
I_{\textrm{IEC}}^\prime &=\int{\mathbf{m}_2\times\mathbf{H}_{\textrm{IEC},2}}\textrm{dV}\\
&=\frac{2A_{12}}{\mu_0M_2\varepsilon}\int{\mathbf{m}_2\times\mathbf{m}_1}\textrm{dV}\\
&=\xi\frac{2A_{12}}{\mu_0M_2}\mathbf{m}_2\times\mathbf{m}_1.
\end{split}
\end{equation}
Uniting Eq. (6)-(8), it leads to
\begin{equation}
\label{eq.9}
{A_2\mathbf{m}}_2\times\frac{\partial\mathbf{m}_2}{\partial y}={A_{12}\mathbf{m}}_1\times\mathbf{m}_2.
\end{equation}
Eq. (5) and Eq. (9) are the complete boundary conditions, and thus can be rewritten together as \cite{RN28,RN33,RN43}
\begin{equation}
\label{eq.10}
\left\{
\begin{aligned}
{A_1\mathbf{m}}_1\times\frac{\partial\mathbf{m}_1}{\partial y}&={A_{12}\mathbf{m}}_1\times\mathbf{m}_2\\
{A_2\mathbf{m}}_2\times\frac{\partial\mathbf{m}_2}{\partial y}&={A_{12}\mathbf{m}}_1\times\mathbf{m}_2.
\end{aligned}
\right.
\end{equation}
A small fluctuation of $\textbf{m}_{\textrm{n}}$ with $\textbf{m}_{\textrm{n}} = \textbf{m}_{\textrm{0,n}} + \widetilde{\textrm{\textbf{m}}}_{\textrm{n}}$ is assumed, where $\textbf{m}_{0,1} = +\textbf{\textrm{e}}_{z}$, $\textbf{m}_{0,2} = +\textbf{\textrm{e}}_{z}$ in the P state and $\textbf{m}_{0,1} = +\textbf{\textrm{e}}_{z}$, $\textbf{m}_{0,2} = -\textbf{\textrm{e}}_{z}$ in the AP state, $\widetilde{\textrm{\textbf{m}}}_{\textrm{n}} = (m_{x,\textrm{n}}, m_{y,\textrm{n}}, 0)$ and $|\textbf{m}_{\textrm{n}}| \ll 1$. In the P case, replacing $\textbf{m}_{\textrm{n}}$ with $\textbf{m}_{\textrm{0,n}} + \widetilde{\textrm{\textbf{m}}}_{\textrm{n}}$ in Eq. (10) and keeping the linear terms of $\widetilde{\textrm{\textbf{m}}}_{\textrm{n}}$ leads to the boundary conditions \cite{RN43}
\begin{equation}
\label{eq.11}
\left\{
\begin{split}
A_1\frac{\partial{\widetilde{\mathbf{m}}}_1}{\partial y}+A_{12}({\widetilde{\mathbf{m}}}_1-{\widetilde{\mathbf{m}}}_2)&=0\\
A_2\frac{\partial{\widetilde{\mathbf{m}}}_2}{\partial y}+A_{12}({\widetilde{\mathbf{m}}}_1-{\widetilde{\mathbf{m}}}_2)&=0.
\end{split}
\right.
\end{equation}
In the AP case, the boundary conditions can be written as
\begin{equation}
\label{eq.12}
\left\{
\begin{split}
A_1\frac{\partial{\widetilde{\mathbf{m}}}_1}{\partial y}-A_{12}({\widetilde{\mathbf{m}}}_1+{\widetilde{\mathbf{m}}}_2)&=0\\
{-A}_2\frac{\partial{\widetilde{\mathbf{m}}}_2}{\partial y}-A_{12}({\widetilde{\mathbf{m}}}_1+{\widetilde{\mathbf{m}}}_2)&=0.
\end{split}
\right.
\end{equation}
Considering a negligible damping $\alpha$ and defining a spin-wave function $\psi_{\textrm{n}}(x,t)=m_{x,\textrm{n}}(x,t)-im_{y,\textrm{n}}(x,t)$, the LLG Eq. (1) can be recasted into an effective Schr\"{o}dinger equation \cite{RN30,RN38,RN52,RN53,RN54,RN55,RN56,RN57}
\begin{equation}
\label{eq.13}
i\hbar\frac{\partial\psi_\textrm{n}}{\partial t}=\emph{H}_\textrm{n}\psi_\textrm{n}=\left(\frac{{\hat{p}}^2}{2m_\textrm{n}^\ast}+V_\textrm{n}\right)\psi_\textrm{n},
\end{equation}
where $\hat{p} = -i\hbar\nabla$ is the momentum operator and $m_{\textrm{n}}^{*}=\hbar M_{\textrm{n}}/4\gamma m_{z,\textrm{n}}A_{\textrm{n}}=\hbar M_{\textrm{n}}/4\gamma\sigma_{\textrm{n}}A_{\textrm{n}}$ is the effective mass of magnons. $V_{\textrm{n}}=2\gamma\hbar\sigma_{\textrm{n}}K_{\textrm{n}}/M_{\textrm{n}}$ represents the potential energy. From Eq. (13), the spin-wave dispersion relation can be obtained \cite{RN34}:
\begin{equation}
\label{eq.14}
\begin{split}
\omega_\textrm{n}&=E_\textrm{n}/\hbar\\
&=(\frac{{p}_\textrm{n}^2}{2m_\textrm{n}^\ast}+V_\textrm{n})/\hbar\\
&=\frac{\hbar}{2m_\textrm{n}^\ast}k_\textrm{n}^2+2\gamma{\sigma_\textrm{n}K}_\textrm{n}/M_\textrm{n},
\end{split}
\end{equation}
where $k_{\textrm{n}}^{2}=\textbf{\textit{k}}_{\textrm{n}} \cdot \textbf{\textit{k}}_{\textrm{n}}$, which shows the magnitude of wave vectors. The sign of $\omega_{\textrm{n}}$ is obviously positive (negative) as $\sigma_{\textrm{n}}=+1(-1)$, indicating that the spin-wave polarization is RH (LH).

For simplicity, a normalized wave function is defined as $\hat{\psi}_{\textrm{n}}(x,t)=\psi_{\textrm{n}}(x,t)/|\psi_{\textrm{1i}}(x,t)|$, where $|\psi_{\textrm{1i}}(x,t)|$ represents the amplitude of the incident waves in medium 1, ensuring that $|\hat{\psi}_{\textrm{n}}(x,t)|=1$. Then the space and time part of $\hat{\psi}_{\textrm{n}}(x,t)$ are separated via $\hat{\psi}_{\textrm{n}}(x,t)=\phi_{\textrm{n}}(x)e^{-i \omega t}$.

From Eq. (11) and Eq. (12), the boundary conditions of $\phi_{\textrm{n}}(x)$ can be obtained and written as
\begin{equation}
\label{eq.15}
\left\{
\begin{split}
A_1\frac{\partial\phi_1}{\partial y}+A_{12}(\phi_1-\phi_2)&=0\\
A_2\frac{\partial\phi_2}{\partial y}+A_{12}(\phi_1-\phi_2)&=0
\end{split}
\right.
\end{equation}
and
\begin{equation}
\label{eq.16}
\left\{
\begin{split}
A_1\frac{\partial\phi_1}{\partial y}-A_{12}\left(\phi_1+\phi_2\right)&=0\\
{-A}_2\frac{\partial\phi_2}{\partial y}-A_{12}\left(\phi_1+\phi_2\right)&=0,
\end{split}
\right.
\end{equation}
corresponding to the P and AP cases, respectively.

The following assumptions can be made that $\phi_{1}(x)=\phi_{\textrm{1i}}(x)+\phi_{\textrm{1r}}(x)=e^{i(k_{x}^{\textrm{i}}x+k_{y}^{\textrm{i}}y)}+re^{i(k_{x}^{\textrm{r}}x+k_{y}^{\textrm{r}}y)}$ and $\phi_{2}(x)=\phi_{\textrm{2t}}(x)=te^{i(k_{x}^{\textrm{t}}x+k_{y}^{\textrm{t}}y)}$ with \textit{r} and \textit{t} being the reflection and transmission coefficients. $\textbf{\textit{k}}^{\textrm{i}}=(k_{x}^{\textrm{i}},k_{y}^{\textrm{i}})$, $\textbf{\textit{k}}^{\textrm{r}}=(k_{x}^{\textrm{r}},k_{y}^{\textrm{r}})$ and $\textbf{\textit{k}}^{\textrm{t}}=(k_{x}^{\textrm{t}},k_{y}^{\textrm{t}})$ are the wave vectors of incident, reflected and transmitted waves, respectively. Thus, the relations $(\textbf{\textit{k}}^{\textrm{i}})^{2}=(\textbf{\textit{k}}^{\textrm{r}})^{2}=(k_{1})^{2}$ and $(\textbf{\textit{k}}^{\textrm{t}})^{2}=(k_{2})^{2}$ can be obtained. The tangential component of the wave vector is conserved due to the translational symmetry along the interface \cite{RN38,RN58,RN59}, $k_{x}^{\textrm{i}}=k_{x}^{\textrm{r}}=k_{x}^{\textrm{t}}$. According to the aforementioned analysis, a significant relation is obtained, $k_{y}^{\textrm{i}}=-k_{y}^{\textrm{r}}$. Hence, the reflected angle $\theta_{\textrm{r}}$ must be equal to the incident angle $\theta_{\textrm{i}}$, which is called the law of reflection. In the P state, via Eq. (15), the reflection and transmission coefficients can be derived as
\begin{equation}
\label{eq.17}
\begin{split}
r_\textrm{P}=\frac{A_1A_2k_y^\textrm{i}k_y^\textrm{t}+{iA}_1A_{12}k_y^\textrm{i}-{iA}_2A_{12}k_y^\textrm{t}}{A_1A_2k_y^\textrm{i}k_y^\textrm{t}+{iA}_1A_{12}k_y^\textrm{i}{+iA}_2A_{12}k_y^\textrm{t}}
\end{split}
\end{equation}
and
\begin{equation}
\label{eq.18}
\begin{split}
t_\textrm{P}=\frac{{2iA}_1A_{12}k_y^\textrm{i}}{A_1A_2k_y^\textrm{i}k_y^\textrm{t}+{iA}_1A_{12}k_y^\textrm{i}{+iA}_2A_{12}k_y^\textrm{t}}.
\end{split}
\end{equation}
In the uniform medium, generally, $r_{\textrm{P}}^{2} + t_{\textrm{P}}^{2}=1$ \cite{RN44}. However, in the present system of two different media, $r_{\textrm{P}}^{2} + t_{\textrm{P}}^{2} \neq 1$, due to the amplitude of reflected waves $r_{\textrm{P}}$ being normalized to that of incident waves in medium 1, but not $t_{\textrm{P}}$. Thus, the reflectance and transmittance are defined as $R = r_{\textrm{P}}^{2}$ and $T = 1-r_{\textrm{P}}^{2} =1-R$, respectively. Notably, $t_{\textrm{P}}$ represents the amplitude of transmitted waves. A detailed discussion of \textit{R} and \textit{T} is provided in Section \uppercase\expandafter{\romannumeral3}, A. \textit{R}, \textit{T} $\in$ (0,1) and the refraction of spin waves occurs when the incident angle is less than critical angle ($\theta_{\textrm{i}} < \theta_{\textrm{c}}$). As $\theta_{\textrm{i}}$ reaches or exceeds $\theta_{\textrm{c}}$ ($\theta_{\textrm{i}} \geq \theta_{\textrm{c}}$), $R \equiv 1$ and $T \equiv 0$, the total reflection occurs in medium 1 and the transmitted waves become evanescent in medium 2.

In the AP state, via Eq. (16), the following equations can be obtained:
\begin{equation}
\label{eq.19}
\begin{split}
r_{\textrm{AP}}=\frac{A_1A_2k_y^\textrm{i}k_y^\textrm{t}-{iA}_1A_{12}k_y^\textrm{i}+{iA}_2A_{12}k_y^\textrm{t}}{A_1A_2k_y^\textrm{i}k_y^\textrm{t}-{iA}_1A_{12}k_y^\textrm{i}{-iA}_2A_{12}k_y^\textrm{t}}
\end{split}
\end{equation}
and
\begin{equation}
\label{eq.20}
\begin{split}
t_{\textrm{AP}}=\frac{{2iA}_1A_{12}k_y^\textrm{i}}{A_1A_2k_y^\textrm{i}k_y^\textrm{t}-{iA}_1A_{12}k_y^\textrm{i}{-iA}_2A_{12}k_y^\textrm{t}}.
\end{split}
\end{equation}
Here, $k_{y}^{\textrm{t}}$ is a pure imaginary number, according to Eq. (14). $R = r_{\textrm{AP}}^{2} \equiv 1$ and $T = 1-R \equiv 0$, indicating that total reflection invariably occurs, similar to the magnon blocking effect in magnon junctions \cite{RN60}. Here, $\phi_{\textrm{2t}}(x)=t_{\textrm{AP}}e^{-|k_{y}^{\textrm{t}}|y}e^{ik_{x}^{\textrm{t}}x}=t_{\textrm{AP}}e^{-\frac{y}{L_{12}}}e^{ik_{x}^{\textrm{t}}x}$ is an evanescent wave along $y$ axis with
\begin{equation}
\label{eq.21}
\begin{split}
L_{12}=1/\left|k_y^\textrm{t}\ \right|\
\end{split}
\end{equation}
being the decay length as spin waves propagate from medium 1 to 2. A detailed discussion of spin-wave transmission in AP state is provided in Section \uppercase\expandafter{\romannumeral3}, B.

\begin{figure}[b]
  \centering
  \includegraphics[width=8cm]{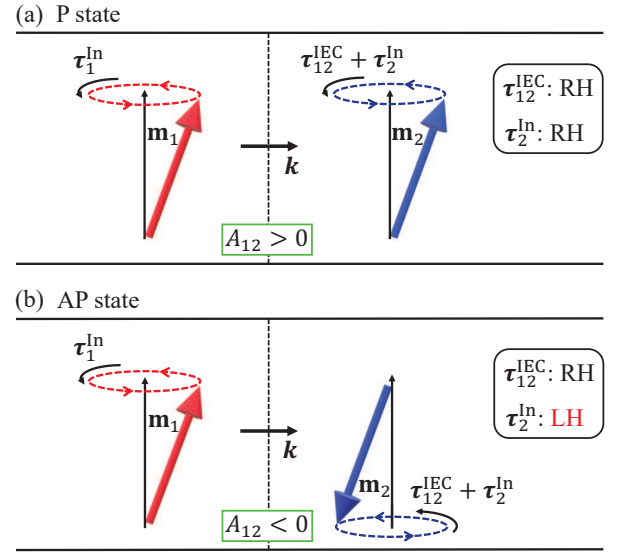}
  \caption{The spin dynamics at the interface in the P and AP states. (a) The precession details of \textbf{m}$_{1}$ (red) and \textbf{m}$_{2}$ (blue) at the interface between two magnetic media with the P configuration, where the type of IEC is ferromagnetic ($A_{12} > 0$). (b) shows the case of the AP configuration, where the IEC is antiferromagnetic ($A_{12} < 0$). The vector $\textbf{\textit{k}}$ represents the spin-wave propagation direction.}\label{Fig2}
\end{figure}

In order to provide a brief understanding of the main difference between the P and AP states, the precession of magnetization near the interface is shown in Figure 2(a) and Figure 2(b). In the P state, $\textbf{m}_{1}$ applies a torque $\bm{\tau}_{12}^{\textrm{IEC}}=-\gamma\mu_{0}\textbf{m}_{2}\times\textbf{\textrm{H}}_{\textrm{IEC,2}}$ on $\textbf{m}_{2}$ due to the IEC. The spin-wave polarization caused by $\bm{\tau}_{12}^{\textrm{IEC}}$ is RH and the same as that caused by $\bm{\tau}_{2}^{\textrm{In}}$, leading to a plane wave $\phi_{\textrm{2t}}(x)=t_{\textrm{P}}e^{i(k_{x}^{\textrm{t}}x+k_{y}^{\textrm{t}}y)}$ in medium 2. By comparison, in the AP state, the spin-wave polarization caused by $\bm{\tau}_{12}^{\textrm{IEC}}$ is still RH but that caused by $\bm{\tau}_{2}^{\textrm{In}}$ is LH \cite{RN34,RN74}. For the magnetization far away from the interface, $\bm{\tau}_{12}^{\textrm{IEC}}$ disappears and $\bm{\tau}_{2}^{\textrm{In}}$ will be the resistance of precession, resulting in an evanescent wave $\phi_{\textrm{2t}}(x)=t_{\textrm{AP}}e^{-\frac{y}{L_{12}}}e^{ik_{x}^{\textrm{t}}x}$ along the $y$ direction. Therefore, the spin-wave transmission properties are considerably different in the P and AP states.

\section{Results and Discussion}
As discussed in this section, YIG and GdIG are chosen as medium 1 and 2, respectively. The magnetic parameters used in the calculation are as follows: the saturation magnetization, exchange constants and uniaxial magnetic anisotropy constants are $M_{1}=1.5\times10^{5}$ A/m, $A_{1}=3.6\times10^{-12}$ J/m, $K_{1}=10$ J/m$^{3}$ for YIG \cite{RN61,RN62,RN63}, and $M_{2}=0.3\times10^{5}$ A/m, $A_{2}=3\times10^{-12}$ J/m, $K_{2}=4000$ J/m$^{3}$ for GdIG \cite{RN64,RN65,RN66}. The constant of IEC between YIG and GdIG is $A_{12}=3.3\times10^{-3}$ J/m$^{2}$ for the P configuration and $A_{12}=-3.3\times10^{-3}$ J/m$^{2}$ for the AP configuration \cite{RN35,RN67}.
\subsection{P configuration}
\begin{figure}[b]
  \centering
  \includegraphics[width=8.6cm]{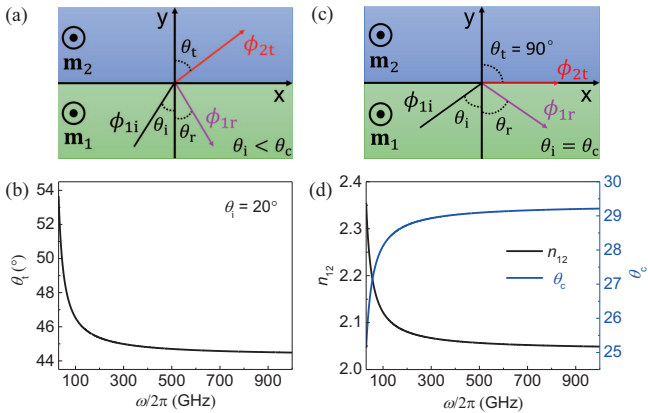}
  \caption{The spin-wave transmission and reflection at the interface between YIG (\textbf{m}$_{1}$) and GdIG (\textbf{m}$_{2}$) with the P configuration. (a) When the incident angle is less than critical angle ($\theta_{\textrm{i}} < \theta_{\textrm{c}}$), spin waves are refracted and reflected. The frequency-dependent refracted angle $\theta_{\textrm{t}}$ is shown in (b) at $\theta_{\textrm{i}}=20^{\circ}$. (c) As $\theta_{\textrm{i}}$ is increased to $\theta_{\textrm{c}}$, spin waves are all reflected. (d) shows the frequency dependence of magnetic refractive index $n_{12}$ (black) and critical angle $\theta_{\textrm{c}}$ (blue).}\label{Fig3}
\end{figure}
A two-medium system is first considered in the P configuration, including YIG (\textbf{m}$_{1}$) and GdIG (\textbf{m}$_{2}$), as shown in Figure 3(a). The frequency $\omega/2\pi$ ranges from 30 to 1000 GHz to ensure that spin waves are exchange-dominated in the system, which has been strictly demonstrated in previous research by the present authors \cite{RN34}. Reflection and refraction occur when the incident angle is less than the critical angle ($\theta_{\textrm{i}} < \theta_{\textrm{c}}$). The reflected angle $\theta_{\textrm{r}}$ is invariably equal to the incident angle $\theta_{\textrm{i}}$, due to the law of reflection as aforementioned. The refracted angle $\theta_{\textrm{t}}$ can be obtained by the relation
\begin{equation}
\label{eq.22}
\begin{split}
k_1\sin{\left(\theta_\textrm{i}\right)}=k_2\sin{\left(\theta_\textrm{t}\right)},
\end{split}
\end{equation}
which is the law of refraction for spin waves, or referred to as the magnonic Snell's law \cite{RN27,RN28,RN29,RN30,RN31,RN32,RN33}. Figure 3(b) is the $\theta_{\textrm{t}}$-$\omega/2\pi$ curve at $\theta_{\textrm{i}}=20^{\circ}$. With the increase of $\omega/2\pi$, $\theta_{\textrm{t}}$ decreases, indicating that the spin-wave path is deflected more in the lower frequency range. As $\theta_{\textrm{t}}$ is increased to $90^{\circ}$, there are no refracted waves and spin waves are all reflected back, as shown in Figure 3(c). Thus, the critical angle $\theta_{\textrm{c}}$ can be obtained by $k_{1}\sin(\theta_{\textrm{c}})=k_{2}\sin(90^{\circ})$, leading to
\begin{equation}
\label{eq.23}
\begin{split}
\theta_\textrm{c}=\arcsin{\left(\frac{k_2}{k_1}\right)}.
\end{split}
\end{equation}
To characterize the extent of refraction, a magnetic refractive index can be defined as
\begin{equation}
\label{eq.24}
\begin{split}
n_{12}=1/\sin{\left(\theta_\textrm{c}\right)}=\frac{k_1}{k_2}.
\end{split}
\end{equation}
Figure 3(d) shows the magnetic refractive index $n_{12}$ (black) and critical angle $\theta_{\textrm{c}}$ (blue) as a function of $\omega/2\pi$, demonstrating that as the frequency becomes higher, the critical angle becomes larger and the magnetic refractive index becomes smaller.

The \textit{T}-$\omega/2\pi$ curves are calculated at different $\theta_{\textrm{i}}$, as shown in Figure 4. When the incident waves are normal to the interface ($\theta_{\textrm{i}}=0^{\circ}$), the transmittance \textit{T} as a function of frequency $\omega/2\pi$ is shown in Figure 4(a). Notably, the transmission spectra depend on the value of $A_{12}$, which is discussed in the Supplemental Material \cite{SM}. In Figure 4(a)-(c), as $\theta_{\textrm{i}}$ is increased, the transmittance \textit{T} decreases, indicating that spin waves are more likely to pass through the interface with the smaller incident angle. As $\theta_{\textrm{i}}$ is further increased to $28^{\circ}$ and $29^{\circ}$ in Figure 4(d) and (e), the transmission spectra are divided into two intervals. The total reflection occurs in the low-frequency interval due to the increased $n_{12}$ in such a range, while the high-frequency spin waves can propagate into GdIG. In Figure 4(f), as $\theta_{\textrm{i}}$ is increased to $30^{\circ}$, the transmittance \textit{T} is zero, which means that all the spin waves in 30-1000 GHz range are totally reflected.

\begin{figure}[b]
  \centering
  \includegraphics[width=8.6cm]{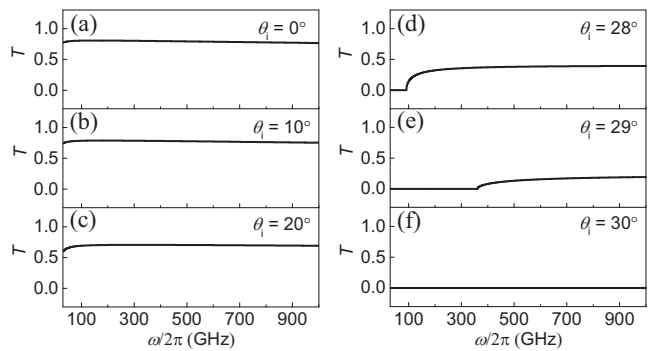}
  \caption{The transmission spectra of spin waves in YIG (\textbf{m}$_{1}$)/GdIG (\textbf{m}$_{2}$) heterojunction with the P configuration. (a)-(f) show the cases of $\theta_{\textrm{i}}=0^{\circ}$, $10^{\circ}$, $20^{\circ}$, $28^{\circ}$, $29^{\circ}$, and $30^{\circ}$.}\label{Fig4}
\end{figure}

In the case of $\theta_{\textrm{i}} > \theta_{\textrm{c}}$, the transmitted spin waves are no longer plane waves, but evanescent waves. The analogous phenomenon has been extensively investigated in optics \cite{RN70,RN71,RN72,RN73}, wherein the light waves turn into evanescent waves when $\theta_{\textrm{i}} > \theta_{\textrm{c}}$. In the present two-medium system, according to $(\textbf{\textit{k}}^{\textrm{i}})^{2}=(k_{1})^{2}$, $(\textbf{\textit{k}}^{\textrm{t}})^{2}=(k_{2})^{2}$, $k_{x}^{\textrm{i}}=k_{x}^{\textrm{t}}$ as well as $\theta_{\textrm{i}} > \theta_{\textrm{c}}$, a conclusion can be drawn that $k_{y}^{\textrm{t}}$ is a pure imaginary number. As a result, $\phi_{\textrm{2t}}(x)=\phi_{\textrm{2t}}^{\textrm{e}}(x)=t_{\textrm{P}}e^{-|k_{y}^{\textrm{t}}|y}e^{ik_{x}^{\textrm{t}}x}=t_{\textrm{P}}e^{-\frac{y}{L_{12}}}e^{ik_{x}^{\textrm{t}}x}$, which is an evanescent wave along $y$ axis with $L_{12}=1/|k_{y}^{\textrm{t}}|$ representing the decay length of spin waves propagating from YIG (\textbf{m}$_{1}$) to GdIG (\textbf{m}$_{2}$). In Figure 5(a), the red dashed line represents the evanescent waves. YIG (\textbf{m}$_{1}$) and GdIG (\textbf{m}$_{2}$) are semi-infinite, and thus the evanescent waves will decay to infinity after departing from the interface. The calculation result of the decay length $L_{12}$ is shown in Figure 5(b). $L_{12}$ obviously depends on the incident angle $\theta_{\textrm{i}}$ and spin-wave frequency $\omega/2\pi$. The minimum of $\theta_{\textrm{i}}$ is $30^{\circ}$, ensuring $\theta_{\textrm{i}} > \theta_{\textrm{c}}$ in 30-1000 GHz. As $\theta_{\textrm{i}}$ is increased from $30^{\circ}$ to the vicinity of $90^{\circ}$, $L_{12}$ decreases monotonically. $L_{12}$ characterizes the decay length along $y$ instead of $x$ axis. As a consequence, spin waves will decay more quickly in the $y$ direction if $\textbf{\textit{k}}^{t}$ is inclined to the $x$ axis as $\theta_{\textrm{i}}$ is increased. The frequency dependence of $L_{12}$ is also monotonic. The data reveal that spin waves tend to have longer decay lengths in the low-frequency range. As $\omega/2\pi$ is increased, the evanescent waves are more concentrated at the interface. This monotonicity is similar to the skin effect of electromagnetic waves in the air-metal system, which can can be referred to as the magnonic skin effect (MSE) \cite{RN34}. Notably, the MSE was first reported in YIG/GdIG heterojunction with the AP configuration, where the MSE is applicable for $\theta_{\textrm{i}} \in [0^{\circ},90^{\circ})$ without the limit of the critical angle $\theta_{\textrm{c}}$. By contrast, the MSE occurs at $\theta_{\textrm{i}} \in [\theta_{\textrm{c}},90^{\circ})$ in the P case.

\begin{figure}[b]
  \centering
  \includegraphics[width=8.6cm]{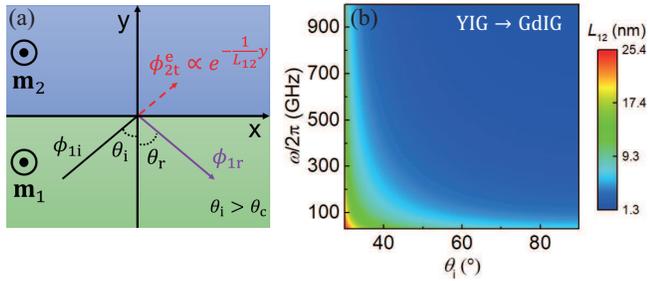}
  \caption{The decay lengths of evanescent waves in the case of $\theta_{\textrm{i}} > \theta_{\textrm{c}}$. (a) As the incident angle $\theta_{\textrm{i}}$ exceeds the critical angle $\theta_{\textrm{c}}$, spin waves are all reflected back. The transmitted waves are evanescent, which have the form of $e^{-\frac{y}{L_{12}}}$ in the $y$ direction with $L_{12}$ being the decay length as spin waves propagate from YIG (\textbf{m}$_{1}$) to GdIG (\textbf{m}$_{2}$). (b) shows the calculation result of the decay length $L_{12}$, which is dependent of the incident angle and frequency.}\label{Fig5}
\end{figure}

Here, an observation can be made that the characteristics of spin-wave transmission in the P and AP configurations are not always different. The distinction and connection depend on the scope of $\theta_{\textrm{i}}$. For the AP state, spin waves are all reflected and only the evanescent waves penetrate through the interface as $\theta_{\textrm{i}} \in [0^{\circ},90^{\circ})$. For the P state, when $\theta_{\textrm{i}} \in [0^{\circ},\theta_{\textrm{c}})$, spin waves can propagate into GdIG (\textbf{m}$_{2}$) in the form of plane waves and are partly reflected back to YIG (\textbf{m}$_{1}$). When $\theta_{\textrm{i}} \in (\theta_{\textrm{c}},90^{\circ})$, total reflection occurs and the spin waves penetrate through the interface in the form of decay, which are the same as those in the AP state. Attention should be paid to the value of $k_{y}^{\textrm{t}}$, which is central to understanding the aforementioned phenomena. In the AP state, $k_{y}^{\textrm{t}}$ is a pure imaginary number, causing the total reflection to be $R = r_{\textrm{P}}^{2} \equiv 1$ according to Eq. (19). For the small $\theta_{\textrm{i}}$ cases in the P state, $k_{y}^{\textrm{t}}$ is a real number, showing the different manifestation compared with the AP state. As $\theta_{\textrm{i}}$ overtakes $\theta_{\textrm{c}}$, $k_{y}^{\textrm{t}}$ turns into a pure imaginary number, resulting in resemblance to the AP case.

\begin{figure}[b]
  \centering
  \includegraphics[width=8.6cm]{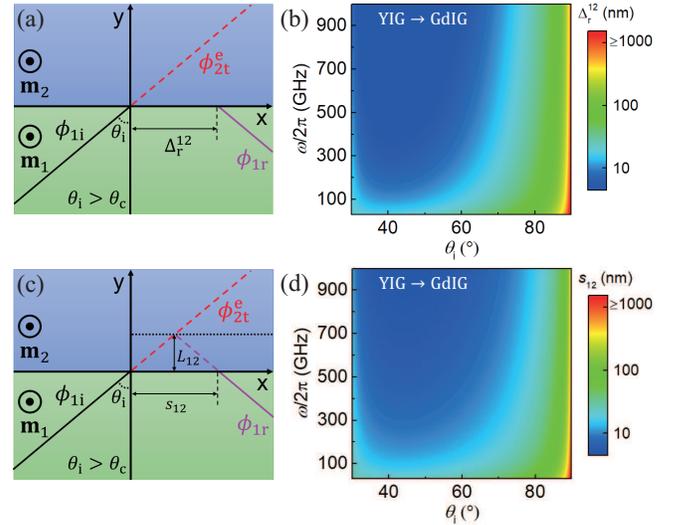}
  \caption{The GH effect at the interface in the case of $\theta_{\textrm{i}} > \theta_{\textrm{c}}$. (a) shows the GH shift of the reflected waves, marked with $\Delta_{r}^{12}$. (b) is the result of $\Delta_{r}^{12}$ calculated by Eq. (25), which depends on $\theta_{\textrm{i}}$ and $\omega/2\pi$. (c) shows the relationship between the lateral shift $s_{12}$ and decay length $L_{12}$. The black dot line represents the effective reflecting interface. (d) The result of $s_{12}$ calculated by Eq. (26). The color bars in (b) and (d) are shown in the log scale.}\label{Fig6}
\end{figure}

In general, the GH effect occurs when spin waves are scattered at the interface \cite{RN35,RN36,RN37,RN38}, describing the shift between the reflected (or transmitted) point and the incident point. According to previous studies, the sign of GH shifts can be positive or negative, depending on the reflection (or transmission) coefficients. As $\theta_{\textrm{i}} > \theta_{\textrm{c}}$, the GH shift of the totally reflected spin waves can be calculated by \cite{RN35,RN36,RN37,RN38}
\begin{equation}
\label{eq.25}
\begin{split}
\Delta_\textrm{r}^{12}=-\frac{\partial\varphi_\textrm{r}}{\partial k_x^\textrm{i}},
\end{split}
\end{equation}
where $\varphi_{r}=\arctan[\textrm{Im}(r_{\textrm{P}})/\textrm{Re}(r_{\textrm{P}})]$ is the phase difference between the reflected and incident waves. $\textrm{Re}(r_{\textrm{P}})$ and $\textrm{Im}(r_{\textrm{P}})$ are the real and imaginary parts of the reflection coefficient $r_{\textrm{P}}$, respectively. Figure 6(a) shows the schematic of the GH shift $\Delta_{r}^{12}$ when spin waves propagate from YIG (\textbf{m}$_{1}$) to GdIG (\textbf{m}$_{2}$). Figure 6(b) shows the computed result of $\Delta_{r}^{12}$ as a function of $\theta_{\textrm{i}}$ and $\omega/2\pi$. An observation can be made that all the values of $\Delta_{r}^{12}$ are positive, ranging from several nanometers to micron scales. The frequency dependence of $\Delta_{r}^{12}$ is monotonic, indicating that high-frequency waves tend to have smaller GH shifts. Although the relation between $\Delta_{r}^{12}$ and $\theta_{\textrm{i}}$ is non-monotonic, $\Delta_{r}^{12}$ increases as $\theta_{\textrm{i}}$ is increased in most of the region. In particular, $\Delta_{r}^{12}$ goes to infinity as $\theta_{\textrm{i}}$ is close to $90^{\circ}$, showing a divergence and being consistent with the results of previous studies \cite{RN35,RN38}. Hence, for the design of spin-wave devices, the GH shift must be considered if $\theta_{\textrm{i}}$ is large.

The underlying physics of the positive GH shift should be considered beyond the mathematical calculation of Eq. (25). The positive GH shift is assumed to originate from the displacement of the effective reflecting interface, as shown in Figure 6(c). The reflection does not occur as soon as the incident waves reach the $x$ axis. A possible scenario is that spin waves penetrate into GdIG in the form of decay and are reflected at another effective interface, resulting in a positive shift $s_{12}$ of the reflected point along $x$ axis. The effective reflecting interface is represented by the black dot line and the displacement is assumed as the decay length $L_{12}$. The following equation is evident:
\begin{equation}
\label{eq.26}
\begin{split}
s_{12}=2L_{12}\tan{\left(\theta_\textrm{i}\right)}.
\end{split}
\end{equation}
In order to verify reasonability of the assumption, the computed result of $s_{12}$ is shown in Figure 6(d), serving as a contrast to $\Delta_{r}^{12}$ in Figure 6(b). The color bars are shown in the log scale due to the divergence in the vicinity of $\theta_{\textrm{i}}=90^{\circ}$. An observation can be made that there is a slight difference in the magnitude of $s_{12}$ and $\Delta_{r}^{12}$. Despite such difference, the incident-angle and frequency dependence are consistent. The comparability demonstrates that the present assumption is reasonable to some extent. Thus, conclusion can be drawn that the positive GH shift can be attributed to the penetration of the evanescent waves in GdIG, and the decay length is roughly equal to the displacement of the effective reflecting interface. Notably, $\Delta_{r}^{12}$ is a function of $A_{12}$, but $s_{12}$ is independent of $A_{12}$. $s_{12}$ and $\Delta_{r}^{12}$ are approximately equal, since $A_{12}$ is relatively large. If $A_{12}$ tends to zero, $\Delta_{r}^{12}$ is also decreased to zero. The influence of $A_{12}$ on the GH shift $\Delta_{r}^{12}$ is shown in the Supplemental Material \cite{SM}. Such findings demonstrate that $\Delta_{r}^{12}$ can be explained by $s_{12}$ when the IEC between YIG and GdIG is strong.

Beyond investigating the scattering problems at a single interface in the two-medium systems, in the following, the spin-wave scattering at multiple interfaces in the multi-medium systems is explored. The YIG (\textbf{m}$_{1}$)/GdIG (\textbf{m}$_{2}$)/YIG (\textbf{m}$_{3}$) heterostructure is first considered, as shown in Figure 7(a) with $\theta_{\textrm{i}} < \theta_{\textrm{c}}$ and Figure 7(e) with $\theta_{\textrm{i}} > \theta_{\textrm{c}}$. In the case of $\theta_{\textrm{i}} < \theta_{\textrm{c}}$, the incident waves are refracted and reflected at two interfaces. Via the transfer matrix method (TMM) \cite{RN44,RN68,RN69}, the transmission coefficients can be obtained. The transmittance \textit{T} as a function of frequency $\omega/2\pi$ is shown in Figure 7(b)-(d) at different $\theta_{\textrm{i}}$ when $\theta_{\textrm{i}} < \theta_{\textrm{c}}$. The results reveal that spin waves can totally pass through GdIG (\textbf{m}$_{2}$) at a special frequency, that is, $T=1$, and such a phenomenon is referred to as the spin-wave resonant transmission effect. As $\theta_{\textrm{i}}$ increases, the position of the resonant peak moves to a higher frequency. The resonant transmission is general in systems with two interfaces \cite{RN42,RN43,RN44,RN45}, which is closely related to the coherence of spin waves. In the present system, when the phases of $\phi_{\textrm{2i}}$ and $\phi_{\textrm{2r}}$ meet certain conditions, resonant transmission occurs.

\begin{figure}[t]
  \centering
  \includegraphics[width=8.6cm]{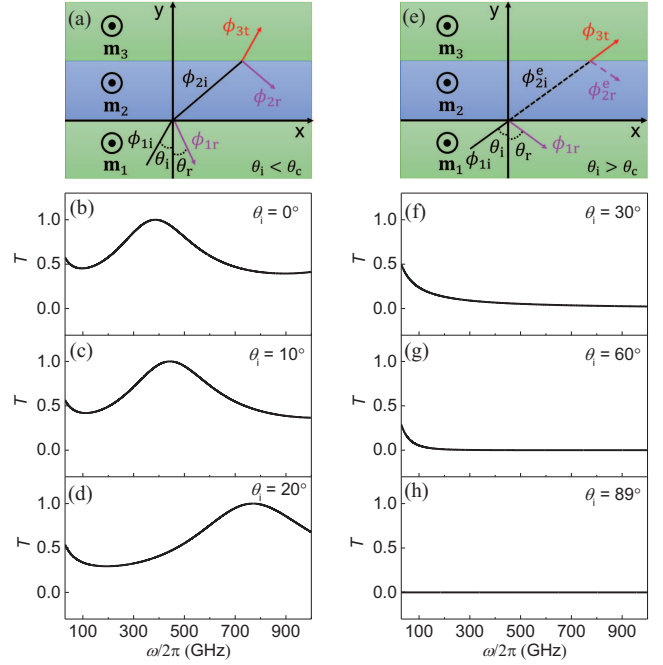}
  \caption{The transmission spectra of spin waves in YIG (\textbf{m}$_{1}$)/GdIG (\textbf{m}$_{2}$)/YIG (\textbf{m}$_{3}$) with the P configuration. (a)-(d) and (e)-(h) correspond to the cases of $\theta_{\textrm{i}} < \theta_{\textrm{c}}$ and $\theta_{\textrm{i}} > \theta_{\textrm{c}}$, respectively. (a) In the case of $\theta_{\textrm{i}} < \theta_{\textrm{c}}$, the spin waves are refracted and reflected at two interfaces. (b)-(d) show the spin-wave resonant transmission effect at $\theta_{\textrm{i}}=0^{\circ}$, $10^{\circ}$ and $20^{\circ}$, where the resonant peak is moved to higher frequency as $\theta_{\textrm{i}}$ increases. (e) As $\theta_{\textrm{i}} > \theta_{\textrm{c}}$, spin waves in GdIG (\textbf{m}$_{2}$) become evanescent and then propagate into YIG (\textbf{m}$_{3}$) in the form of plane waves, which is called the spin-wave tunnelling. The transmission spectra at $\theta_{\textrm{i}}=30^{\circ}$, $60^{\circ}$ and $89^{\circ}$ are shown in (f)-(h). The thickness of GdIG (\textbf{m}$_{2}$) is 10 nm.}\label{Fig7}
\end{figure}
\begin{figure}[t]
  \centering
  \includegraphics[width=8.6cm]{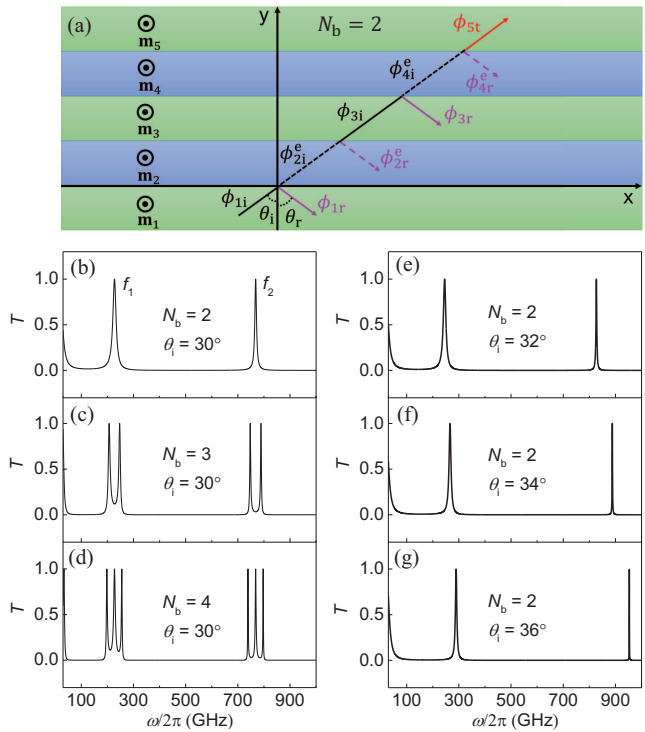}
  \caption{The spin-wave resonant tunnelling effect when $\theta_{\textrm{i}} > \theta_{\textrm{c}}$. (a) The spin-wave propagation in YIG (\textbf{m}$_{1}$)/GdIG (\textbf{m}$_{2}$)/YIG (\textbf{m}$_{3}$)/GdIG (\textbf{m}$_{4}$)/YIG (\textbf{m}$_{5}$) with the P configuration. The medium 2 and medium 4 are barriers and thus spin waves therein are evanescent. $N_{\textrm{b}}$ represents the number of GdIG barriers. (b)-(d) are the transmission spectra at $\theta_{\textrm{i}}=30^{\circ}$ with $N_{\textrm{b}}=2$, 3 and 4. The number of resonant tunnelling peaks increase as $N_{\textrm{b}}$ is increased. (e)-(g) are the transmission spectra at $N_{\textrm{b}}=2$ with $\theta_{\textrm{i}}=32^{\circ}$, $34^{\circ}$ and $36^{\circ}$. The resonant tunnelling frequency increases as $\theta_{\textrm{i}}$ is increased. The thickness of GdIG (\textbf{m}$_{2}$), YIG (\textbf{m}$_{3}$) and GdIG (\textbf{m}$_{4}$) are 10, 10 and 10 nm.}\label{Fig8}
\end{figure}

In Figure 7(e), $\phi_{\textrm{2i}}^{\textrm{e}}$ and $\phi_{\textrm{2r}}^{\textrm{e}}$ are evanescent waves with no coherence, thus the resonant transmission does not exist. Moreover, the total reflection is also nonexistent. The evanescent waves decay along $y$ axis in GdIG (\textbf{m}$_{2}$) and turn into the plane waves $\phi_{\textrm{3t}}$ in YIG (\textbf{m}$_{3}$). Such a phenomenon is referred to as spin-wave tunnelling due to the transmission being by means of the evanescent waves. Figure 7(f)-(h) show the frequency-dependent transmittance \textit{T} at $\theta_{\textrm{i}}=30^{\circ}$, $60^{\circ}$ and $89^{\circ}$. As the frequency $\omega/2\pi$ or incident angle $\theta_{\textrm{i}}$ is increased, the decay length $L_{12}$ decreases, thus the transmittance diminishes. As $\omega/2\pi$ is high enough or $\theta_{\textrm{i}}$ is close to $90^{\circ}$, the decay length tends to zero, leading to the total reflection.

As the number of media is further increased to five at $\theta_{\textrm{i}} > \theta_{\textrm{c}}$ as shown in Figure 8(a), another resonance effect arises, which is referred to as spin-wave resonant tunnelling effect. The difference between the resonant transmission and resonant tunnelling can be described as follows. The resonant tunnelling is a special case of the resonant transmission. If the evanescent waves are involved in the transmission process, the phenomenon of $T=1$ is referred to as the resonant tunnelling \cite{RN40,RN41}. In Figure 8(b), two resonant tunnelling peaks with resonant frequency $f_{1}$ and $f_{2}$ at $N_{\textrm{b}}=2$ and $\theta_{\textrm{i}}=30^{\circ}$ can be observed, where $N_{\textrm{b}}$ is the number of GdIG. For spin waves, GdIG (\textbf{m}$_{2}$) and GdIG (\textbf{m}$_{4}$) can be regarded as the potential barriers, where the plane waves are forbidden. Thus $N_{\textrm{b}}$ also represents the number of barriers. As $N_{\textrm{b}}$ is increased to three in Figure 8(c), the single peak at $f_{1}$ (or $f_{2}$) splits into two. In Figure 8(d), $N_{\textrm{b}}=4$, the single peak becomes three. The results indicate that the number of resonant tunnelling peaks depends on the number of barriers. As $N_{\textrm{b}}$ increases, the single peak will split into $N_{\textrm{b}}-1$. Figure 8(b) and (e)-(g) exhibit the influence of $\theta_{\textrm{i}}$ on the resonant tunnelling. Apparently, the peaks are moved to high frequency as the incident angle $\theta_{\textrm{i}}$ increases. Moreover, in Figure 8, an observation can be made that the full width at half maximum (FWHM) of the high-frequency peaks tends to be narrower. The narrow FWHM is useful in the design of the spin-wave filter, which only allows spin waves with a particular frequency to transmit.

The spin-wave transmission in multi-medium systems shows significant differences with the two-medium systems. Such differences are closely related to the number of the scattering interfaces. In two-medium systems, there is no resonance effect due to the single interface. Hence, the primary condition of resonance is the existence of at least two scattering interfaces. Meanwhile, the existence of plane waves are also essential between the two interfaces. Otherwise, like the case in Figure 7(e), the resonance cannot occur.

\subsection{AP configuration}
In this section, a discussion is provided on the spin-wave transmission in the systems with the AP configuration. The YIG (\textbf{m}$_{1}$)/GdIG (\textbf{m}$_{2}$) heterojunction with antiferromagnetic IEC is considered, as shown in Figure 9(a). Due to the inverse spin-wave polarization, the spin waves are all reflected back and propagate into GdIG (\textbf{m}$_{2}$) in the form of evanescent waves $\phi_{\textrm{2t}}^{\textrm{e}}$. Based on such a mechanism, the effects of the total reflection and decay are expected to be independent of materials, and only depend on the polarization inversion of the two magnetic media. To test such a claim, the direction of the spin-wave propagation is inverted, as shown in Figure 9(c). On the basis of Eq. (19), the reflectance $R = r_{\textrm{AP}}^{2} \equiv 1$ can be demonstrated, including whether the direction is from YIG (\textbf{m}$_{1}$) to GdIG (\textbf{m}$_{2}$), or from GdIG (\textbf{m}$_{2}$) to YIG (\textbf{m}$_{1}$), invariably leading to the total reflection. Such inverse-polarization-induced total reflection has been presented by the previous works \cite{RN34,RN74}. Additionally, the results of the decay lengths $L_{12}=1/|k_{2y}^{\textrm{12}}|$ and $L_{21}=1/|k_{1y}^{\textrm{21}}|$ are given in Figure 9(b) and Figure 9(d), corresponding to the cases of Figure 9(a) and Figure 9(c), respectively. The incident-angle and frequency dependence of decay lengths are similar to those shown in Figure 5(b). With the increase of $\theta_{\textrm{i}}$, both $L_{12}$ and $L_{21}$ decrease. As $\omega/2\pi$ is increased, the decay lengths decrease rapidly, referred to as the MSE, resembling the case of $\theta_{\textrm{i}} > \theta_{\textrm{c}}$ in the aforementioned P state. Notably, the MSE in the YIG/GdIG heterojunction with the AP configuration has been investigated in the previous research by the present author through both theoretical method and micromagnetic simulation \cite{RN34}. Further, an obvious difference of the decay length $\Delta L$ can be found at given $\omega/2\pi$ and $\theta_{\textrm{i}}$ between the two paths. The formula of $\Delta L$ can be written as
\begin{equation}
\label{eq.27}
\begin{split}
\Delta L=L_{21}-L_{12}=\frac{1}{\left|k_{1y}^{21}\ \right|}-\frac{1}{\left|k_{2y}^{12}\right|}\\
=\frac{\left|k_{2y}^{12}\right|^2-\left|k_{1y}^{21}\ \right|^2}{\left|k_{1y}^{21}\ \right|\left|k_{2y}^{12}\right|\left(\left|k_{2y}^{12}\right|+\left|k_{1y}^{21}\ \right|\right)}.
\end{split}
\end{equation}
The denominator is positive, and the numerator is defined as $\Delta |k|^{2}=|k_{2y}^{\textrm{12}}|^{2}-|k_{1y}^{\textrm{21}}|^{2}$. According to Eq. (14), the following can be obtained:
\begin{equation}
\label{eq.28}
\begin{split}
\Delta{|k|}^2=\frac{\omega}{2\gamma}\left(\frac{M_1}{A_1}-\frac{M_2}{A_2}\right)\cos^2(\theta_{\textrm{i}})\\
+\left(\frac{K_1}{A_1}-\frac{K_2}{A_2}\right)\left[1+\sin^2(\theta_{\textrm{i}})\right]
\end{split}
\end{equation}
Thus, the conditions of $\Delta |k|^{2}=0$ ($\Delta L=0$) are $\frac{M_{1}}{A_{1}}-\frac{M_{2}}{A_{2}}=0$ and $\frac{K_{1}}{A_{1}}-\frac{K_{2}}{A_{2}}=0$. In other words, $m_{1}^{*}+m_{2}^{*}=0$ and $V_{1}+V_{2}=0$. Obviously, $\Delta L \neq 0$ because $m_{1}^{*}+m_{2}^{*} \neq 0$ and $V_{1}+V_{2} \neq 0$ in the system, showing that the nonreciprocity of the decay lengths results from the asymmetries of the effective mass of magnons and potential energy of two media.

\begin{figure}[t]
  \centering
  \includegraphics[width=8.6cm]{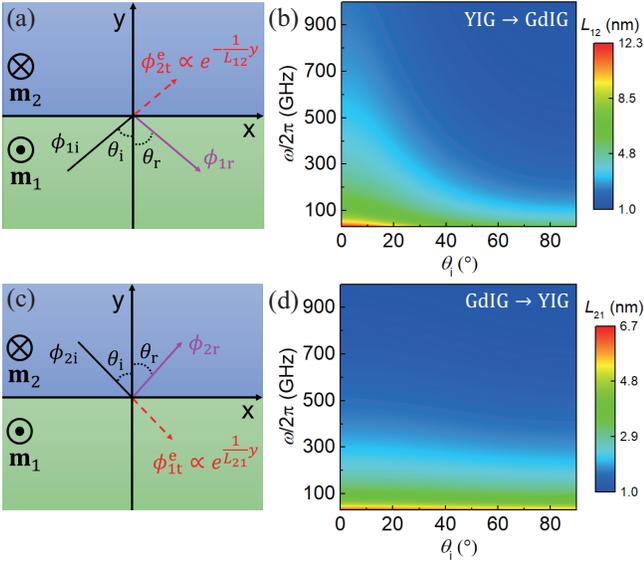}
  \caption{The decay lengths of evanescent waves in YIG (\textbf{m}$_{1}$)/GdIG (\textbf{m}$_{2}$) heterojunction with the AP configuration. (a) Spin waves propagate from YIG (\textbf{m}$_{1}$) to GdIG (\textbf{m}$_{2}$). The waves in GdIG (\textbf{m}$_{2}$) have the form of $e^{-\frac{y}{L_{12}}}$, which are evanescent due to the reversed spin-wave polarization compared with YIG (\textbf{m}$_{1}$). (c) shows the case that spin waves propagate from GdIG (\textbf{m}$_{2}$) to YIG (\textbf{m}$_{1}$). The evanescent waves in YIG (\textbf{m}$_{1}$) is in the form of $e^{\frac{y}{L_{21}}}$. (b) and (d) are the calculation results of decay lengths, corresponding to the cases of (a) and (c), respectively.}\label{Fig9}
\end{figure}
\begin{figure}[t]
  \centering
  \includegraphics[width=8.6cm]{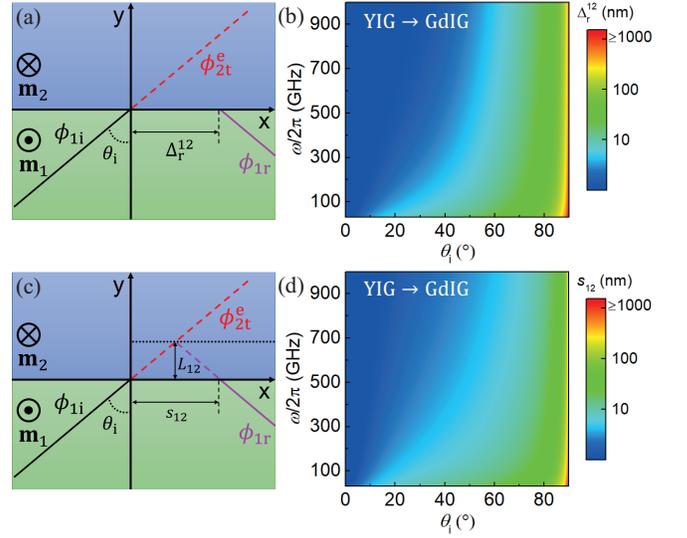}
  \caption{The GH effect in the AP configuration as spin waves propagate from YIG (\textbf{m}$_{1}$) to GdIG (\textbf{m}$_{2}$). (a) shows the GH shift of the reflected waves, represented by $\Delta_{r}^{12}$. (b) is the computed result of $\Delta_{r}^{12}$. (c) The relationship between the lateral shift $s_{12}$ and the decay length $L_{12}$. The black dot line represents the effective reflecting interface. (d) is the computed result of $s_{12}$. The colar bars in (b) and (d) are shown in the log scale.}\label{Fig10}
\end{figure}
\begin{figure*}[t]
  \centering
  \includegraphics[width=14cm]{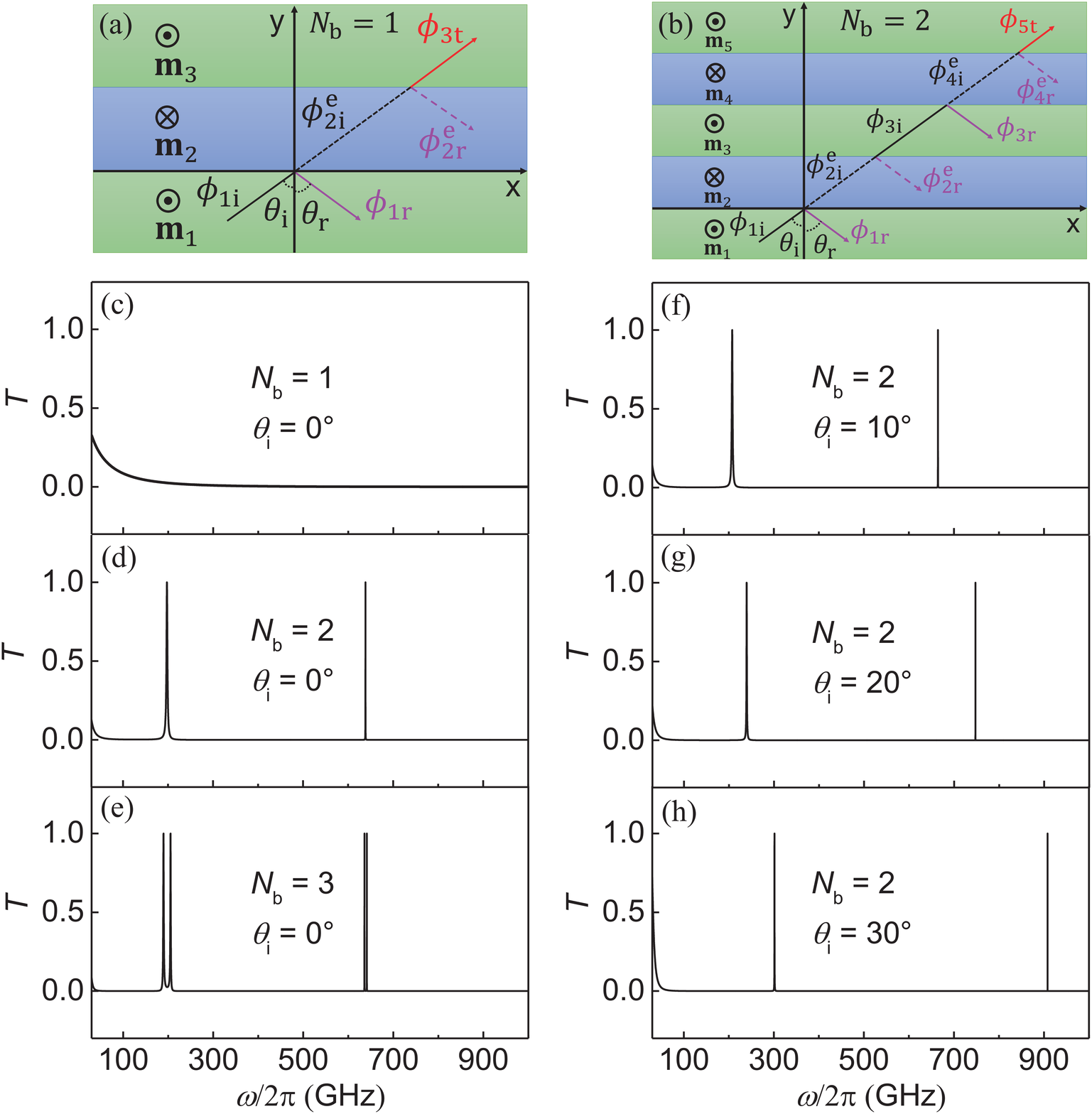}
  \caption{The spin-wave tunnelling and resonant tunnelling effect in the multi-medium system with the AP configuration. (a) The spin-wave tunnelling in YIG (\textbf{m}$_{1}$)/GdIG (\textbf{m}$_{2}$)/YIG (\textbf{m}$_{3}$), which is the single barrier structure ($N_{\textrm{b}}=1$). The thickness of GdIG (\textbf{m}$_{2}$) is 10 nm. (b) is the $N_{\textrm{b}}=2$ case, corresponding to the structure YIG (\textbf{m}$_{1}$)/GdIG (\textbf{m}$_{2}$)/YIG (\textbf{m}$_{3}$)/GdIG (\textbf{m}$_{4}$)/YIG (\textbf{m}$_{5}$), where the spin-wave resonant tunnelling occurs. (c)-(e) are the transmission spectra at $\theta_{\textrm{i}}=0^{\circ}$ with $N_{\textrm{b}}=1$, 2 and 3. (f)-(h) are the transmission spectra at $N_{\textrm{b}}=2$ with $\theta_{\textrm{i}}=10^{\circ}$, $20^{\circ}$ and $30^{\circ}$. The thickness of GdIG (\textbf{m}$_{2}$), YIG (\textbf{m}$_{3}$) and GdIG (\textbf{m}$_{4}$) are 10, 10 and 10 nm.}\label{Fig11}
\end{figure*}

Figure 10(a) shows the GH shift $\Delta_{r}^{12}$ of the reflected spin waves. $\Delta_{r}^{12}$ is calculated by the equation $\Delta_\textrm{r}^{12}=-\frac{\partial\varphi_\textrm{r}}{\partial k_x^\textrm{i}}$, where $\varphi_{r}=\arctan[\textrm{Im}(r_{\textrm{AP}})/\textrm{Re}(r_{\textrm{AP}})]$ is the phase difference between the reflected and incident waves with $\textrm{Re}(r_{\textrm{AP}})$ and $\textrm{Im}(r_{\textrm{AP}})$ being the real and imaginary parts of the reflection coefficient $r_{\textrm{AP}}$, respectively. The computed results are shown in Figure 10(b). As $\theta_{\textrm{i}}$ increases, or $\omega/2\pi$ decreases, $\Delta_{r}^{12}$ becomes larger monotonically. In particular, when $\theta_{\textrm{i}}$ approaches $90^{\circ}$, $\Delta_{r}^{12}$ tends to infinity. As such, in order to obtain the precise spin-wave path, the GH shift must be taken into consideration.

The shift $s_{12}$ is subsequently calculated by the equation $s_{12}=2L_{12}\tan{\left(\theta_\textrm{i}\right)}$ to investigate the connection with the GH shift $\Delta_{r}^{12}$. Figure 10(c) shows the relationship between $s_{12}$ and the decay length $L_{12}$. The spin waves are not reflected at $x$ axis, but propagate into GdIG in the form of decay and are reflected at another effective interface marked by the black dot line. Such a mechanism contributes to the positive lateral shift $s_{12}$. The computed results of $s_{12}$ are given in Figure 10(d). By comparing Figure 10(b) and Figure 10(d), an observation can be made that the incident-angle and frequency dependence of $\Delta_{r}^{12}$ and $s_{12}$ are the same, despite a slight difference in magnitude. At the same time, the distinction of $\Delta_{r}^{12}$ and $s_{12}$ in magnitudes indicates that the exact reflecting interface is not the black dot line, but in the vicinity of said line. Thus, the decay length is approximately equal to the real penetration depth of spin waves in GdIG. Similar to the P state, the similarity between $\Delta_{r}^{12}$ and $s_{12}$ is applicable only in the case that the IEC between YIG and GdIG is strong. Besides, when spin waves propagate from GdIG to YIG, the results of $\Delta_{r}^{12}$ and $s_{21}$ are analogous to Figure 10, which are shown in the Supplemental Material \cite{SM}.

The spin-wave transmission in multi-medium systems with the AP configuration is also investigated. Figure 11(a) and (b) show the YIG (\textbf{m}$_{1}$)/GdIG (\textbf{m}$_{2}$)/YIG (\textbf{m}$_{3}$) and YIG (\textbf{m}$_{1}$)/GdIG (\textbf{m}$_{2}$)/YIG (\textbf{m}$_{3}$)/GdIG (\textbf{m}$_{4}$)/YIG (\textbf{m}$_{5}$) heterostructures, corresponding to the number of GdIG barriers $N_{\textrm{b}}=1$ and $N_{\textrm{b}}=2$, respectively. Spin waves propagate in the form of evanescent waves in GdIG, but they are plane waves in YIG, due to the inverse spin-wave polarization. Figure 11(c) is the computed result in the case of $N_{\textrm{b}}=1$ and $\theta_{\textrm{i}}=0^{\circ}$. The nonzero transmittance \textit{T} demonstrates that spin waves indeed pass through the GdIG barrier. As the frequency is increased, the decay length becomes shorter in GdIG, and thus the transmittance decreases monotonically. For $N_{\textrm{b}}=2$, the transmission spectra are shown in Figure 11(d), where the resonant tunnelling effect appears. Spin waves can fully transmit across the two GdIG barriers at the resonant frequency. For $N_{\textrm{b}}=3$, as shown in Figure 11(e), the resonant peaks split. The number of splitting is $N_{\textrm{b}}-1$ for every single peak as $N_{\textrm{b}}>2$. Figure 11(f)-(h) shows the influence of $\theta_{\textrm{i}}$ on the resonant tunnelling. As $\theta_{\textrm{i}}$ is increased, the resonant peaks are moved to high-frequency ranges. In addition, the FWHM of the resonant peaks in Figure 11 are much narrower than those of the P configuration in Figure 8.

By comparing the spin-wave characteristics in P and AP configurations, the differences and similarities can be observed. In the P configuration, the critical angle $\theta_{\textrm{c}}$ is vital. For $\theta_{\textrm{i}} < \theta_{\textrm{c}}$, there is no decay process of spin waves in all of the media. The refraction and reflection are the key phenomena here. Moreover, if the number of scattering interfaces is increased to two or more, resonant transmission occurs. Contrastingly, in the case of $\theta_{\textrm{i}} > \theta_{\textrm{c}}$, evanescent waves appear, leading to entirely different features, such as total reflection, decay, positive GH shift, tunnelling and resonant tunnelling effect. In the AP configuration, spin waves show considerably similar characteristics with the $\theta_{\textrm{i}} > \theta_{\textrm{c}}$ case of the P configuration. Although the aforementioned phenomena are various and complex, several traces can be found, as well as connections therebetween.

\section{Conclusion}
In summary, a theoretical method was developed for the transmission of spin waves with nanoscale wavelengths in the P and AP magnetic configurations. Through the method, the transmission and reflection coefficients were both analytically and numerically investigated. The computed results show several phenomena, including spin-wave refraction, total reflection, decay, positive GH shifts, tunneling, resonant transmission and resonant tunneling. When the incident angle is smaller than the critical angle, the spin-wave polarization is a significant factor in the different performances of the two configurations, demonstrating that the same polarization leads to transmission while the inverse polarization results in total reflection. Such effect is similar to the spin-dependent scattering in GMR \cite{RN2,RN3}, TMR \cite{RN1,RN9,RN10} and MVE \cite{RN19,RN20}. Hence, in the present study, the spin-wave scattering process in P ($\theta_{\textrm{i}} < \theta_{\textrm{c}}$) and AP configurations can be referred to as polarization-dependent scattering. As the incident angle exceeds the critical angle, the situation in the P configuration is significantly changed and shows similarities with the AP case. The transmitted spin waves turn into evanescent waves, of which the decay lengths were explored and a connection with the positive GH shifts was found. Moreover, the investigation of spin waves propagating in multi-medium systems displays the resonance effect, which is closely related to the spin-wave coherence. The present study can facilitate further understanding of the transmission of spin waves in different magnetic configurations and can provide guidance for the design of future magnonic devices.

\begin{acknowledgments}
We thank T. Y. Zhang for helpful discussions. This work was supported by the National Key Research and Development Program of China [MOST, Grants No. 2017YFA0206200 and 2021YFB3201801], the National Natural Science Foundation of China [NSFC, Grants No.51831012 and No.51620105004], Beijing Natural Science Foundation (Grant No. Z201100004220006) and partially supported by the Strategic Priority Research Program (B) [Grant No. XDB33000000] of the Chinese Academy of Sciences (CAS).
\end{acknowledgments}

\bibliographystyle{apsrev4-2}

%

\end{document}